\documentclass[sigconf]{acmart}

\usepackage{algorithmic}
\usepackage{graphicx}
\usepackage{textcomp}

\usepackage{enumitem}
\usepackage{todonotes}
\usepackage{printlen}

\usepackage{url}
\usepackage{booktabs}
\usepackage{caption}
\usepackage{multirow}
\usepackage{censor}
\usepackage{hyperref}
\usepackage{soul}

\usepackage{subcaption}
\usepackage{mathrsfs}
\usepackage{numprint}
\npdecimalsign{,}
\nprounddigits{2}

\makeatletter
\newcommand{\FontInfo}{%
  \texttt{Font=\fontname\font\quad Size=\f@size pt}%
}
\makeatother
\definecolor{main}{HTML}{5989cf}    
\definecolor{sub}{HTML}{cde4ff}     

\definecolor{yel}{rgb}{1.0000,0.6902,0.0000} 
\definecolor{ora}{rgb}{0.9961,0.3804,0.0000} 
\definecolor{pin}{rgb}{0.8627,0.1490,0.4980} 
\definecolor{pur}{rgb}{0.4706,0.3686,0.9412} 
\definecolor{blu}{rgb}{0.3922,0.5608,1.0000} 

\usepackage[many]{tcolorbox}    	
\usepackage{tikz}
\usetikzlibrary{shapes.geometric, arrows.meta, positioning, fit, backgrounds, calc}

\newtcolorbox{RQanswer}[1][]{%
    colback=blu!50,
    colframe=black!5,
    notitle,
    sharp corners,
    borderline west={2pt}{0pt}{main!80!black},
    enhanced,
    breakable,
    right=6pt,
    top=0pt,
    bottom=0pt,
    }
    
\usepackage{breqn}
\usepackage{amsthm}

\usepackage{tablefootnote}

\AtBeginDocument{%
  }

\newif\ifchanges
\changestrue 
\NewEnviron{changes}{%
  \ifchanges
    \color{blue}\BODY
  \else
    \BODY
  \fi
}

\begin{document}

\title{Model See, Model Do? \\
Exposure-Aware Evaluation of Bug-vs-Fix Preference in Code LLMs}

\author{Ali Al-Kaswan}
\orcid{0000-0001-7338-2044}
\affiliation{%
  \institution{Delft University of Technology}
  \city{Delft}
  \country{Netherlands}
}
\email{a.al-kaswan@tudelft.nl}

\author{Claudio Spiess}
\orcid{0009-0000-5932-7022}
\affiliation{%
  \institution{University of California at Davis}
  \city{Davis}
  \country{United States}
}
\email{cvspiess@ucdavis.edu}

\author{Prem Devanbu}
\orcid{0000-0002-4346-5276}
\affiliation{%
  \institution{University of California at Davis}
  \city{Davis}
  \country{United States}
}
\email{ptdevanbu@ucdavis.edu}

\author{Arie van Deursen}
\orcid{0000-0003-4850-3312}
\affiliation{%
  \institution{Delft University of Technology}
  \city{Delft}
  \country{Netherlands}
}
\email{arie.vandeursen@tudelft.nl}

\author{Maliheh Izadi}
\orcid{0000-0001-5093-5523}
\affiliation{%
  \institution{Delft University of Technology}
  \city{Delft}
  \country{Netherlands}
}
\email{m.izadi@tudelft.nl}


\begin{abstract}
Large language models are increasingly used for code generation and debugging, 
but their outputs can still contain bugs that originate from training data.
Distinguishing whether an LLM prefers correct code,
or a familiar incorrect version
might be influenced by what it's been exposed
to during training. 
We introduce an exposure-aware evaluation framework that quantifies how prior exposure to buggy versus fixed code influences a model's
preference.
Using the ManySStuBs4J benchmark, 
we apply Data Portraits for membership testing on the Stack-V2 corpus to estimate whether each buggy and fixed variant was seen during training. 
We then stratify examples by exposure and compare model preference using code completion as well as multiple likelihood-based scoring metrics.
We find that most examples (67\%) have neither variant in the training data, and when only one is present, fixes are more frequently present than bugs.
In model generations, models reproduce buggy lines far more often than fixes, 
with bug-exposed examples amplifying this tendency and fix-exposed examples showing only marginal improvement. 
In likelihood scoring, minimum and maximum token-probability metrics consistently prefer the fixed code across all conditions, indicating a stable bias toward correct fixes.
In contrast, metrics like the Gini coefficient reverse preference when only the buggy variant was seen.
Our results indicate that exposure can skew bug-fix evaluations
and highlight the risk that LLMs may propagate memorised errors in practice.
\end{abstract}


\maketitle

\section{Introduction}
Large Language Models (LLMs) have revolutionised code-related tasks in software engineering, serving as powerful tools for code completion, generation, and debugging~\cite{hou2023large, izadi2022codefill, xia_agentless_2024, izadi2024language}. Their popularity stems from their ability to understand and produce code in multiple programming languages, often outperforming traditional methods in terms of speed and accessibility~\cite{nam2024using, fakhoury2024llm}. For instance, models like GitHub Copilot, built on top of LLMs, have been widely adopted by developers, with millions of users leveraging them daily for productivity gains~\cite{ziegler2024measuring}. 

Although LLMs have transformed code generation and debugging workflows, their output is not immune from errors; this introduces substantial risks that can undermine software reliability and security~\cite{tambon2025bugs, alkaswan2023abuse}. For example, LLM-generated code has been found to include various kinds of errors~\cite{jesse2023large, mohsin2024can, zhong2024can}. These bugs can propagate through systems, leading to costly failures, data breaches, or exploits in real-world applications. These bugs may stem from coding errors in training data~\cite{tambon2025bugs, jahanshahi2025cracks}. 

The ManySStubs4J dataset collects thousands of SStuBs, (`\textbf{S}imple, \textbf{St}upid \textbf{B}ug\textbf{s}'), which are single-statement bugs and their fixes from open-source Java repositories~\cite{karampatsis2020often}. Although seemingly minor, these SStuBs are particularly impactful because they represent real-world errors, mined from open-source projects. Previous work has shown that LLMs are almost twice as likely to produce a buggy version of a line compared to the fixed version~\cite{jesse2023large}.

A critical aspect exacerbating the risks associated with LLM-generated code is the issue of memorisation. LLMs can memorise and regurgitate parts of their training data~\cite{carlini2021extracting, schwarzschild2024rethinking, kong2025demystifying, meeus2025sok}. This issue is particularly pronounced in code LLMs, where training data often consists of vast, publicly sourced, repositories that contain a mix of correct and erroneous code, leading to potential biases in model outputs~\cite{alkaswan2023abuse, jahanshahi2025cracks}. 

For example, if a model has been exposed to a specific buggy code sequence during training, perhaps due to its prevalence in open source datasets, the model may inadvertently favour and propagate code similar to the buggy version, rather than the fixed version~\cite{jahanshahi2025cracks}; indeed subsequent generations from the model could reflect memorised buggy code rather than correct coding patterns, either learned directly from corrected versions or via appropriate generalizations~\cite{alkaswan2023abuse, alkaswan2024traces, wan2024does}. If memorised code is regurgitated, this can affect evaluation metrics in benchmarks~\cite{wan2024does, alkaswan2023abuse, yang2024unveiling}, which raises concerns about the reliability of LLM-assisted code development~\cite{yang2024unveiling, alkaswan2024traces}.

Beyond correctness, exposure-influenced behaviour of code LLMs creates concrete security and governance problems. From a security perspective, models that reproduce insecure patterns in training data can introduce systemic vulnerabilities and expand the attack surface across many downstream projects~\cite{yang2024unveiling, alkaswan2023abuse, mohsin2024can, jahanshahi2025cracks}; memorisation of proprietary or sensitive snippets also raises legal risks if such content is regurgitated~\cite{alkaswan2023abuse, alkaswan2024traces, katzy2024exploratory, katzy2025heapcontaminationfreemultilingualcode}. Therefore, there is a need for exposure-aware methods that accurately assess how prior exposure exposure to data influences LLMs.

To investigate whether LLMs ``see, then do'' when it comes to buggy versus fixed code, we develop an exposure‑aware evaluation pipeline that samples bug-fix pairs from mined benchmarks, and estimates training exposure for each variant using membership tools. We then measure model preference with a suite of likelihood‑based metrics. Additionally we evaluate model completions based on the bug-fix context. We then relate preference to exposure and to bug categories, and perform robustness checks to identify which metrics and categories are most sensitive to exposure, with the goal of disentangling memorisation‑driven propagation from genuine learned correctness. 

Our evaluation yields several noteworthy findings. First, we find that about 67\% of bug–fix pairs in SStuBs were unseen in the model’s training data, and furthermore, the fixes were substantially more likely to have been seen than the bugs. Second, in actual model generations, buggy variants are reproduced far more often than fixes. Moreover, exposure to the buggy variant substantially boosts bug-matching rates in generations, while exposure to the fix provides only marginal improvements in fix-generation rates. These patterns are consistent across evaluated models. Third, likelihood-based metrics such as the minimum and maximum token probability consistently favour the correct, fixed code across all exposure conditions, whereas others like the Gini coefficient can exhibit exposure-driven reversals (preferring the buggy code when the model has seen that bug in training).  Finally, model preferences vary by bug category, with some robust to exposure, others highly susceptible to propagation, and a few exhibiting intrinsic bug preference.

We make the following contributions:
\begin{description}
    \item[Exposure-Aware Evaluation Methodology:] We introduce a no\-vel framework to assess code LLM behaviour that accounts for training data exposure, we perform membership inference for each SStuBs sample to determine if the buggy or fixed variant appeared in the model’s training data. 
    \item[Insights on Metrics and Generations:] We compare likelihood-based metrics to identify those sensitive to exposure and break down preferences by bug category. 
    \item[Empirical Study of Bug Propagation:] We conduct an extensive study on the SStuBs benchmark across multiple open-source LLMs, quantifying how prior exposure to buggy or fixed code influences model preferences and the propensity to generate bugs.
\end{description}

\section{Background and Related Work}
LLMs have been applied to code completion and repair, but errors in generated code remain a serious concern. Researchers have assembled benchmarks of real bugs to measure repair and detection tools. For example, the ManySStuBs4J dataset collects thousands of single-statement bugs and their fixes from open-source Java repositories~\cite{karampatsis2020often}. Other standard benchmarks include Defects4J~\cite{just2014defects4j} and QuixBugs~\cite{lin2017quixbugs}. which capture small but impactful errors (off-by-one mistakes, incorrect conditionals, missing null checks, etc.) in real programs. Recent studies find that large code models still produce buggy code at non-trivial rates: \citeauthor{jesse2023large} report that an LLM is almost twice as likely to generate a buggy line (from SStuBs) as the correct fix~\cite{jesse2023large} and that bugs introduced by others take much longer to fix~\cite{jesse2023large}. Security analyses similarly show that model-synthesised code can include insecure patterns such as missing input sanitization or unsafe API usage~\cite{tambon2025bugs, mohsin2024can}. 

A major challenge in evaluating such models is memorisation of training data. Code corpora contain many duplicate snippets, and large models are known to memorise frequent patterns~\cite{alkaswan2024traces, hu2022membership, zeng2023exploring}. This raises the classic membership inference problem: given a code snippet, was it in the training set? In practice this is hard to determine, especially in cases where the training data is not available~\cite{carlini2021extracting, shokri2017membership, meeus2025sok}.

Prior analyses have shown that training/test overlap can skew evaluation: test snippets discovered in training data can inflate reported performance~\cite{karmakar2022codex}. For bug–fix pairs, the asymmetry is especially problematic: a bug appears first in a project and may propagate to forks before it is fixed, so the buggy version may be more prevalent in scraped data than the fix. \citeauthor{jahanshahi2025cracks}~\cite{jahanshahi2025cracks} found that 17\% of the blobs in the Stack-v2 are copied and have a newer version of the code available. Of these newer versions, around 17\% are bug fixes. These `orphan vulnerabilities' can persist in public repositories even after a fix has been issued~\cite{jahanshahi2025cracks, reid2022extent}.

Despite these insights, existing work has not explicitly controlled for exposure when comparing models’ behaviour on bugs versus fixes. Previous work notes that models sometimes echo insecure or erroneous patterns from their training data~\cite{mireshghallah2022memorization, hartmann2023sokmemorizationgeneralpurposelarge, yang2024unveiling, alkaswan2024traces, ma2025safety, di2025llms, salerno2025much}. But none of these quantify how differing exposure to the buggy versus fixed code affects model preferences. In effect, previous evaluations may confound a model's learned correctness with simple memorisation. In this work, we address these gaps with an exposure-aware analysis of bug–fix preference.

\section{Approach}
\label{sec:approach}
\begin{figure*}[!t]
  \centering 
  \begin{tikzpicture}[
      node distance=6mm and 12mm,
      box/.style={rectangle, draw, fill=blue!10, minimum width=22mm, minimum height=8mm, align=center, font=\footnotesize},
      data/.style={cylinder, shape border rotate=90, aspect=0.3, draw, fill=ora!30, minimum height=8mm, minimum width=20mm, align=center, font=\footnotesize},
      process/.style={rectangle, draw, fill=pur!30, minimum width=22mm, minimum height=8mm, align=center, font=\footnotesize},
      arrow/.style={->, thick}
    ]
    
    \node[process] (sample) {Sample pairs\\
    \raisebox{-1mm}{\tikz[scale=0.7]{
      \draw[fill=pin!15] (-0.7,-0.3) rectangle (-0.1,0.3);
      \draw[gray!60, thick] (-0.6,0.2) -- (-0.2,0.2);
      \draw[gray!60, thick] (-0.6,0.1) -- (-0.3,0.1);
      \draw[gray!60, thick] (-0.6,0.0) -- (-0.4,0.0);
      \draw[gray!60, thick] (-0.6,-0.1) -- (-0.3,-0.1);
      \draw[pin!70, thick] (-0.6,-0.2) -- (-0.2,-0.2);
      \node at (0,0) {|};  
      \draw[fill=blu!15] (0.1,-0.3) rectangle (0.7,0.3);
      \draw[gray!60, thick] (0.2,0.2) -- (0.6,0.2);
      \draw[gray!60, thick] (0.2,0.1) -- (0.5,0.1);
      \draw[gray!60, thick] (0.2,0.0) -- (0.4,0.0);
      \draw[gray!60, thick] (0.2,-0.1) -- (0.5,-0.1);
      \draw[blu!70, thick] (0.2,-0.2) -- (0.6,-0.2);
    }}};
    
    \node[process, right=of sample] (membership) {Estimate exposure\\
    \raisebox{-1mm}{\includegraphics[height=6mm]{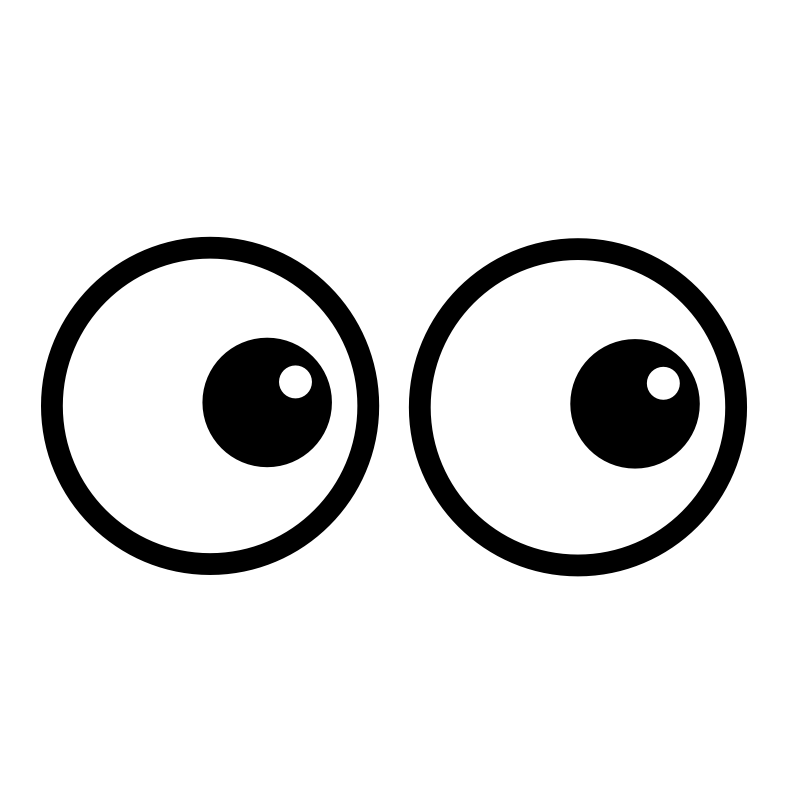}}};
    \node[process, right=of membership] (stratify) {Stratify by exposure\\
    \raisebox{-2mm}{\tikz[scale=0.6]{
      \draw[fill=pin!30] (-0.6,0.0) rectangle (0.0,0.6);
      \fill[pin!70] (-0.5,0.5) circle (0.025);
      \fill[pin!70] (-0.3,0.4) circle (0.025);
      \fill[pin!70] (-0.2,0.2) circle (0.025);
      \fill[pin!70] (-0.4,0.3) circle (0.025);
      \draw[fill=yel!30] (0.0,0.0) rectangle (0.6,0.6);
      \fill[yel!70] (0.1,0.5) circle (0.025);
      \fill[yel!70] (0.3,0.3) circle (0.025);
      \fill[yel!70] (0.5,0.4) circle (0.025);
      \fill[yel!70] (0.2,0.1) circle (0.025);      
      \draw[fill=gray!30] (-0.6,-0.6) rectangle (0.0,0.0);
      \fill[gray!60] (-0.5,-0.2) circle (0.025);
      \fill[gray!60] (-0.3,-0.4) circle (0.025);
      \fill[gray!60] (-0.2,-0.5) circle (0.025);      
      \draw[fill=blu!30] (0.0,-0.6) rectangle (0.6,0.0);
      \fill[blu!70] (0.1,-0.3) circle (0.025);
      \fill[blu!70] (0.3,-0.5) circle (0.025);
      \fill[blu!70] (0.5,-0.2) circle (0.025);
      \fill[blu!70] (0.2,-0.4) circle (0.025);
    }}};
    \node[process, right=of stratify] (scoring) {Model scoring\\
    \raisebox{-1mm}{\tikz[scale=0.6]{
      \draw[fill=yel!60] (0,0) rectangle (0.15,0.4);
      \draw[fill=ora!60] (0.15,0) rectangle (0.3,0.8);
      \draw[fill=pin!60] (0.3,0) rectangle (0.45,0.6);
      \draw[fill=pur!60] (0.45,0) rectangle (0.6,0.3);
      \draw[fill=blu!60] (0.6,0) rectangle (0.75,0.5);
    }}};
    
    \node[process, above=of scoring] (generation) {Model generations\\
    \raisebox{-1mm}{\includegraphics[height=6mm]{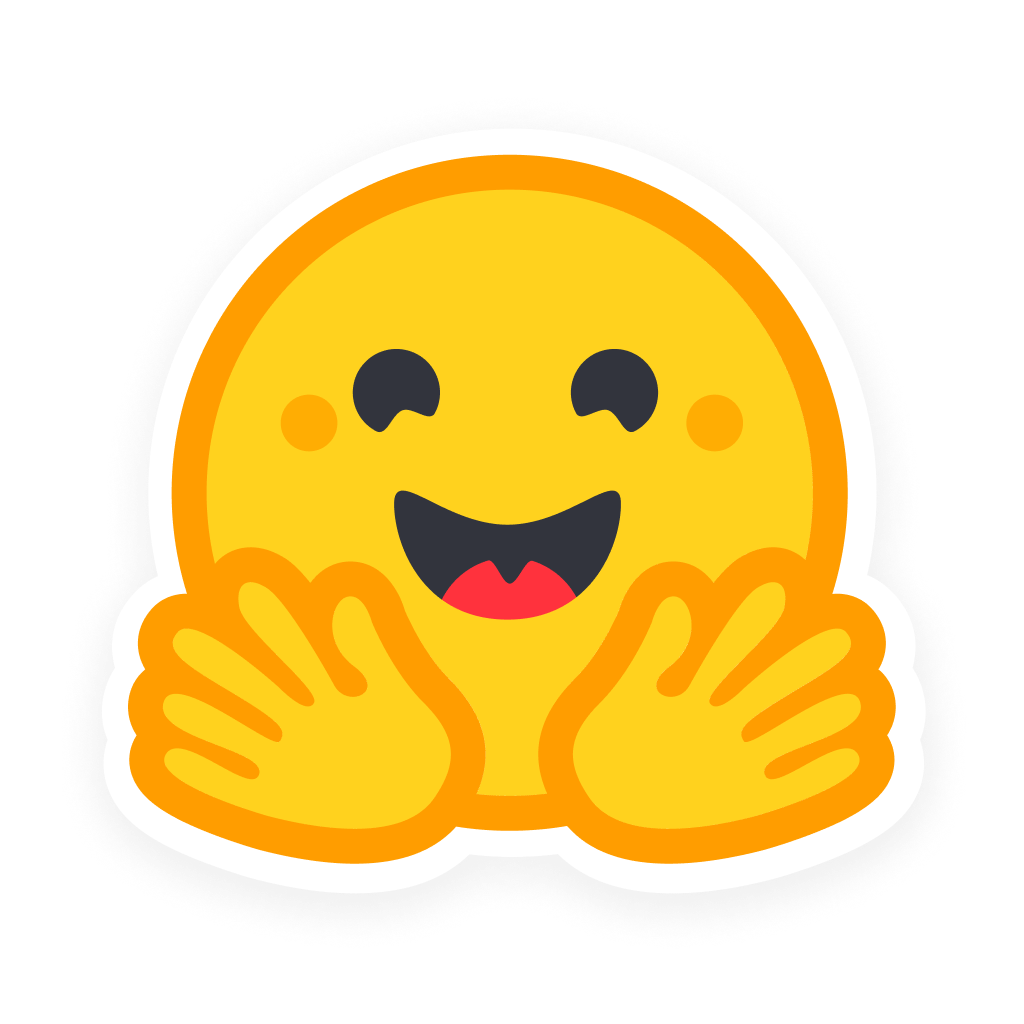}}};
    
    \node[process, right=of generation] (match) {Generation matching\\
    \raisebox{-1mm}{\tikz[scale=0.6]{
      \draw[fill=gray!20, thick] (-1.6,0.15) rectangle (-0.7,0.45);
      \draw[fill=gray!20, thick] (-1.6,0.45) rectangle (-0.7,0.75);
      \draw[gray!40, thick] (-1.5,0.3) -- (-0.8,0.3);
      \draw[gray!40, thick] (-1.5,0.6) -- (-0.8,0.6);
      \draw[fill=ora!20, thick] (-1.6, 0.15) rectangle (-0.7,-0.15);
      \draw[ora!70, thick] (-1.5,0.0) -- (-0.8,0.0);      
      \draw[->, thick] (-0.6,0) -- (0.0,0);    
      \draw[fill=blu!30, thick] (0.1,0.15) rectangle (1.0,0.45);
      \draw[blu!70, very thick] (0.2,0.30) -- (0.9,0.30);
      \draw[fill=pin!20, thick] (0.1,-0.15) rectangle (1.0,0.15);
      \draw[pin!70, thick] (0.2,0.0) -- (0.9,0.0);
      \draw[fill=gray!20, thick] (0.1,-0.15) rectangle (1.0,-0.45);
      \draw[gray!60, thick] (0.2,-0.30) -- (0.9,-0.30);
    }}};
    
    \node[process, below=of match] (analysis) {Analysis \& Results\\
    \raisebox{-1mm}{\tikz[scale=0.4]{
      \draw[fill=blue!20, thick] (0,0) rectangle (0.3,0.8);
      \draw[fill=green!20, thick] (0.4,0) rectangle (0.7,0.5);
      \draw[fill=red!20, thick] (0.8,0) rectangle (1.1,0.3);
      \draw[thick] (-0.1,0) -- (1.2,0);
      \draw[thick] (0,0) -- (0,0.9);
    }}};

    \node[data, above= of sample] (sstubs) {SStuBs\\
    \raisebox{-2mm}{\tikz[scale=0.4]{
      \draw[fill=pin!15, rounded corners=1pt] (-0.9,-0.5) rectangle (-0.1,0.5);
      \draw[gray!60, thick] (-0.8,0.3) -- (-0.2,0.3);    
      \draw[gray!60, thick] (-0.8,0.15) -- (-0.4,0.15);  
      \draw[pin!70, thick] (-0.8,0) -- (-0.3,0);         
      \draw[gray!60, thick] (-0.8,-0.15) -- (-0.5,-0.15); 
      \draw[gray!60, thick] (-0.8,-0.3) -- (-0.6,-0.3);  
      \draw[fill=blu!15, rounded corners=1pt] (0.6,-0.5) rectangle (1.4,0.5);
      \draw[gray!60, thick] (0.7,0.3) -- (1.3,0.3);      
      \draw[gray!60, thick] (0.7,0.15) -- (1.0,0.15);    
      \draw[blu!70, thick] (0.7,0) -- (1.1,0);         
      \draw[gray!60, thick] (0.7,-0.15) -- (1.2,-0.15);  
      \draw[gray!60, thick] (0.7,-0.3) -- (1.1,-0.3);    
    }}};
    \node[data, above= of membership] (stack) {Training data\\
    \raisebox{-2mm}{\tikz{
      \draw[fill=blu!30] (0,0) ellipse (0.35 and 0.05);
      \draw[fill=pur!40] (0,0.08) ellipse (0.35 and 0.05);
      \draw[fill=pin!30] (0,0.16) ellipse (0.35 and 0.05);
      \draw (-0.35,0) -- (-0.35,0.16);
      \draw (0.35,0) -- (0.35,0.16);
    }}};

    \draw[arrow] (sstubs) -- (sample);
    \draw[arrow, dashed] (stack) -- (membership);
    \draw[arrow] (sample) -- (membership);
    \draw[arrow] (membership) -- (stratify);
    \draw[arrow] (stratify) -- (scoring);
    \draw[arrow] (stratify.north) |- (generation.west);
    \draw[arrow] (generation) -- (match);
    \draw[arrow] (match) -- (analysis);
    \draw[arrow] (scoring) -- (analysis);
  \end{tikzpicture}
  \caption{Overview of our exposure-aware evaluation framework}
  \label{fig:approach}
\end{figure*}
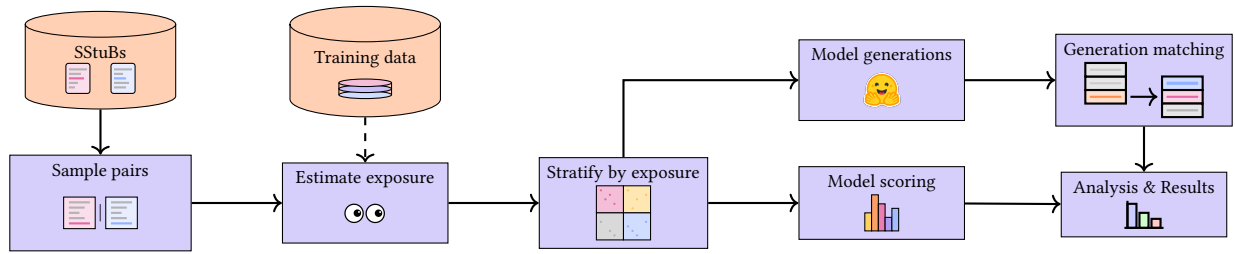

The approach is described in \autoref{fig:approach}. Our exposure-aware evaluation framework studies bug propagation in code LLMs through an end-to-end pipeline: starting from the left, bug-fix pairs are sampled from the SStuBs dataset; exposure is estimated for each variant from the training corpus; samples are stratified into categories; model preferences are evaluated through likelihood-based scoring and actual code generations from LLMs; generations are matched against the original bug or fix variants; and finally, results are analysed to quantify exposure effects on bug-fix preferences.

To study the impact of exposure on bug propagation, we first sample a number of SStuBs from public repos. We then perform membership inference to check if the buggy or fixed version of the code is present in the training data.

With membership inference we can determine whether a bug or its associated fix appears in the large corpus. The samples can be stratified into four sets based on this analysis:
\begin{itemize}
    \item \textbf{Neither} the bug nor the fix are present in the training data
    \item \textbf{Both} the bug and fix are present\footnote{Note that the number of instances data might differ between the bug and fix}
    \item \textbf{Only} the \textbf{bug} is present, but the fix is not
    \item \textbf{Only} the \textbf{fix} is present, but the bug is not 
\end{itemize}

Using a variety of LLMs, we can then calculate the model preference for either the bug or the fix, for each of the previously mentioned sets. We measure both conditional likelihoods and actual generative behaviour. The conditional likelihoods are extracted from the model losses and are scored using different token-level signals. Generation is evaluated using the LLM under decoding.

\subsection{Research Questions}
To guide our evaluation of code LLMs on bug propagation, we address the following research questions (RQs). These build on our hypothesis that training exposure to buggy or fixed code influences model behaviour.

\textbf{RQ1: How does prior exposure to buggy or fixed code influence model preferences as measured by likelihood-based metrics?} By computing metrics such on token probabilities we investigate biases in model scoring, helping disentangle memorisation-driven preferences from learned generalisations.

\textbf{RQ2: How do exposure effects on preferences and generations vary by bug category?} This RQ breaks down results by SStuBs categories to pinpoint which bug types are most susceptible to memorisation.

\textbf{RQ3: How does exposure affect the generation of buggy versus fixed code in model completions?} By focusing on verbatim regurgitation rates, we quantify explicit bug propagation in sampled outputs, assessing real-world risks like vulnerability introduction in downstream applications.

\subsection{Datasets}
Our study utilises two primary datasets: Stack-v2~\cite{lozhkov2024starcoder2stackv2} as the training corpus for membership inference and ManySStuBs4J~\cite{karampatsis2020often} for bug-fix pairs.

To estimate exposure and perform membership inference, we utilise the Stack-v2 dataset. The Stack-v2 is the largest publicly curated code dataset spanning 619 programming languages and totalling 68TB~\cite{lozhkov2024starcoder2stackv2}. The Stack-v2 is mined from the Software Heritage archive~\cite{abramatic2018building}. Its massive size improves the chance of overlaps with the SStuBs data and makes it suitable for training state-of-the-art LLMs. The Stack-v2 is widely used in research as it supports many languages and excludes restrictively licenced  code~\cite{lozhkov2024starcoder2stackv2}.

SStuBs focus on single-statement bugs that can be fixed with single-line changes. SStuBs allow us to isolate changes without the complexity of multi-line or multi-file changes, which allows us to study model behaviour~\cite{karampatsis2020often, just2014defects4j}. We use the ManySStuBs4J dataset, which contains mined bug-fix pairs from open-source Java projects~\cite{karampatsis2020often, jesse2023large}. We opt to use the processed version of the dataset from \citeauthor{jesse2023large} as it contains additional information such as the number of commits between the fix and the bug, as well as the location of the bug in the source file~\cite{jesse2023large}. The dataset consists of $16,899$ bug-fix pairs.

\subsection{Models}
For our evaluation, we focus on models with open weights, and training data where possible, to improve reproducibility. Because of hardware limitations, we omit models that are too expensive to run. Due to hardware limitations we can support models with up to $20B$ parameters. We found that few state-of-the-art models publish their training data. Of the 50 top performing models on the EvalPlus benchmark~\cite{evalplus}, only two have open training data, and both are instruction tuned versions of StarCoder2.\footnote{EvalPlus Benchmark: \url{https://evalplus.github.io/leaderboard.html} (Accessed October 2025)} 

The StarCoder2 family of models is fully open-weight and open-data. The models are based on the Transformer architecture and are available in 3 sizes, 3B, 7B and 15B parameters~\cite{lozhkov2024starcoder2stackv2}.\footnote{StarCoder2: \url{https://huggingface.co/collections/bigcode/starcoder2}} 

StarCoder2 is trained on the Stack-v2 dataset, which is openly available. The 3B and 7B variants are trained on a subset of the Stack-v2 which only includes 17 programming languages (including Java), while the 15B parameter model is trained on the full set of 600+ languages~\cite{lozhkov2024starcoder2stackv2}.

We also include Mellum-4B which is an open-weight (but not fully open-data) model trained on Stack-v2 and some other datasets which are not publicly available~\cite{Mellum-4b-base}.\footnote{Mellum-4B: \url{https://huggingface.co/JetBrains/Mellum-4b-base}} Despite its partially undisclosed training data, we can estimate exposure for SStuBs pairs, as the Stack-v2 forms the core of its code training.

Additionally, we include SmolLM3 (3B)~\cite{bakouch2025smollm3}, a fully open-weight model introduced by Hugging Face in 2025. SmolLM3 was trained on roughly 11 trillion tokens using exclusively public datasets in a staged curriculum that mixes web text, mathematics, and code. The code component of its training is drawn from Stack-v2 and other sources like GitHub issues, notebooks, and Q\&A forums~\cite{bakouch2025smollm3}. We therefore believe it is unlikely that any of the SStuBs ended up in the training data from sources other than Stack-v2. We use the SmolLM3 base checkpoint, without instruction tuning.\footnote{SmolLM3-3B: \url{https://huggingface.co/HuggingFaceTB/SmolLM3-3B-Base}}

\subsection{Data Portraits}
To estimate the exposure of code variants, we utilise Data Portraits~\cite{marone2023data}, a membership inference tool designed to measure the presence of specific text snippets or patterns in large corpora. We use the Data Portrait created by the authors for Stack-v2.\footnote{Stack-v2 Data Portrait: \url{https://stack-v2.dataportraits.org}}

Data Portraits use strided Bloom filters~\cite{bloom1970space, marone2023data} to enable efficient membership testing over large datasets. Bloom filters are probabilistic data structures that can quickly determine whether an element is in a set or not using fixed-length hash functions. Because different strings may map to the same position in the filter, Bloom filters can produce false positives, indicating membership when the element is not actually present, but they do not yield false negatives. Consequently, when applied in Data Portraits, some code snippets flagged as members might not truly exist in the corpus~\cite{marone2023data}.

The false positive rate can be reduced by increasing the number of bits output by the hash function. In the case of Data Portrait these are tuned to achieve a false positive rate of 0.1\%~\cite{marone2023data}. This results in a portrait of 90GB (compared to 68TB for the full dataset) for Stack-v2, which fits in memory and allows for almost instant membership testing~\cite{marone2023data}.

It is important to note a key constraint of this method: its soundness is only guaranteed for queries exceeding a specific length. This results from the construction of the Data Portrait~\cite{marone2023data}. The tool was built using a fixed n-gram \textit{width} (\(w\)) of 50 tokens and a \textit{stride} (\(s\)) of 50 tokens. This means only 50-token sequences starting at every 50-token interval in the training corpus were hashed into the Bloom filter. Consequently, a snippet could fall entirely between these sampling points, leading to a false negative (i.e., the tool reporting a snippet as `unseen' despite its presence in the data). 

To ensure that any query is guaranteed to fully contain at least one of the sampled n-grams, regardless of its alignment, its length (\(L\)) must satisfy the condition \(L \geq s + w - 1\). For the Stack-v2 Data Portrait, with \(s=50\) and \(w=50\), the minimum required query length is \(50 + 50 - 1 = 99\) tokens. Any query of 99 tokens or more is thus guaranteed to contain at least one complete 50-token n-gram that would have been hashed and captured in the Portrait, ensuring the test is sound. We adhere to this constraint for all our membership queries by padding any shorter SStuB to 99 tokens using the surrounding context. In the case of Data Portaits, every character, except spaces and newlines, is considered a token. 

We pad from both sides to ensure that we capture as much of the bug or fix behaviour of the snippet. If a sample is only padded with its preceding context, a hit in the Bloom filter could only cover the context and none of the sample. By padding both sides, we can ensure that the hit covers at least \((99 - sample\_len)/2\) tokens of the sample in the worst case.

When a code snippet is queried, the Data Portrait tool calculates a `badness' score. This score represents the fraction of the snippet's n-grams that are found in the Bloom filter. A higher score indicates a stronger signal that the snippet, or a very similar sequence, is present in the training corpus. For samples with \(sample\_len <= 99\) only a single hit is possible.

A high `badness' score can arise from a query accumulating multiple hits. The cause is the duplicated presence of code blocks in the training corpus. For example, a function may appear in multiple files. Because the stride is applied independently to each file, the function's starting position relative to the beginning of the file dictates which of its internal 50-token n-grams are sampled and hashed. In one file, the stride may capture an n-gram from the beginning of the function, while in another file, a different alignment causes an n-gram from the middle of the same function to be hashed. Consequently, when the full function is used as a query, the sliding-window registers multiple distinct hits against the portrait, one for each independently hashed segment discovered across the corpus~\cite{marone2023data}. Therefore, a single query can generate numerous hits, indicating that different parts of it were seen by the model due to their varied placement across the training data. This underestimates the number of duplicates in the training data as some duplicates are aligned.

To convert this probabilistic score into a binary seen or unseen classification for our study, we apply a threshold. Following the methodology of \citeauthor{marone2023data}, we classify a snippet as exposed (seen) if its badness score is 90\% or higher~\cite{marone2023data}.

We apply Data Portraits to determine whether the SStuBs bug–fix pairs from our dataset were present in Stack-v2, the training corpus used for the models in our study~\cite{lozhkov2024starcoder2stackv2}. 

\subsection{Model Preference Metrics}\label{sec:prefMetrics}
To evaluate the preference of code LLMs between buggy and fixed variants, we compute \(P(w_t \mid w_{t-1})\), the conditional probability of each token comprising the bug or fix, given the preceding tokens i.e., the code lines before the buggy or fixed statement. As the preceding context is the same for any bug-fix pair, it follows that the probabilities are equal. Thus, we consider only the conditional probabilities of the tokens \emph{comprising} the bug and fix code, which may have differing lengths. We collect two token probability sequences, one for the bug and one for the fix, per model and SStuBs pair.

We study multiple likelihood-based metrics to determine model preference based off the token probabilities. The model's preferred variant is selected based on the metric's orientation. For example, if the mean of the token probabilities comprising the fix is higher than the bug, the model prefers the fix. In contrast, a lower perplexity (surprisal) indicates greater preference over a higher perplexity sequence.

However, the choice of metric is not obvious, nor has a singular standard been established in prior work~\cite{shorinwaSurveyUncertaintyQuantification2025}, as metrics capture different aspects of the token probability distribution. Therefore, we evaluate various metrics collected from prior work, which we introduce below.


\begin{description}
    \item[Length \(\uparrow\)] The total number of tokens in the code sequence. While not a direct measure of preference, length can influence probability calculations and is used as a baseline metric~ \cite{spiess2024calibration}.
    \item[Perplexity \(\downarrow\)] A widely used metric in language modelling that measures how well the model predicts a given sequence~\cite{xu2024one}. It is the exponentiation of the average negative log-likelihood per token. Lower perplexity indicates better model confidence and preference for the sequence.
    \item[Minimum Probability \(\uparrow\)] The smallest probability assigned to any token in the sequence. This metric highlights the least confident token prediction, with higher minimum probabilities indicating more uniformly confident predictions.
    \item[Maximum Probability \(\uparrow\)] The highest probability among all tokens in the sequence. This reflects the model's peak confidence at any position in the code snippet, which may indicate familiarity or memorisation.
    \item[Gini Coefficient \(\downarrow\)] A statistical measure of inequality in the distribution of token probabilities, where values closer to zero indicate a more uniform probability distribution, and higher values indicate more skewed confidence towards certain tokens~\cite{bestaReasoningLanguageModels2025}.
    \item[Geometric Mean\(\uparrow\)] The nth root of the product of token probabilities (multiplicative average), which penalises sequences with any low-probability tokens more heavily than high ones, emphasising consistently high or low probabilities across tokens.
    \item[Arithmetic Mean \(\uparrow\)] The arithmetic mean of token probabilities across the sequence. This captures the model's average confidence in the output, can be skewed by outlier high probability tokens. Lower arithmetic mean indicates lower average confidence (higher uncertainty) across the sequence.
\end{description}

\subsection{Model Generations}
To complement the likelihood-based metrics, we evaluate the models' actual generative behaviour. For each bug-fix pair in the SStuBs dataset, we provide the model with the preceding code context (i.e., the code lines before the buggy or fixed statement) as a prompt, limited to 2048 tokens. We then generate completions using \emph{temperature=0.8} and $top\_p=0.95$ following~\citeauthor{lozhkov2024starcoder2stackv2}~\cite{lozhkov2024starcoder2stackv2, gu2024cruxeval}. We generate five independent completions per prompt, to limit the computational cost. Each completion limited to a maximum of $64$ tokens, this is in line with literature~\cite{jesse2023large}. We found that almost all SStuBs were under this limit. 
We then assess whether these generations match the buggy variant, the fixed variant, both (if different generations match each), or neither. Matching is determined by exact string comparison~\cite{jesse2023large}.

We compute the following disjoint aggregate statistics stratified by exposure category, which sum to 100\%:
\begin{description}
    \item[Fix without bugs:] Proportion of samples where at least one generation matches the fixed variant but none match the bug.
    \item[Bug without fixes:] Proportion of samples where at least one generation matches the buggy variant but none match the fix.
    \item[Mix bug-fix:] Proportion of samples where at least one generation matches the bug and at least one matches the fix.
    \item[No bug no fix:] Proportion of samples where no generations match either the buggy or the fixed variant.
\end{description}

\subsection{Environment and Data Availability}
All experiments were run on a server with two Nvidia A40 GPUs with 48GB of VRAM (driver version $535.261.03$), 256GB of RAM and an AMD EPYC 9334 CPU with 64 cores. The experiments were run with Docker. A Dockerfile to reproduce the environment is available in the replication package.

A replication package, including all data and code, is available on Zenodo\footnote{Replication package: \url{https://doi.org/10.5281/zenodo.17423424}}.

\section{Exposure Characterization of Dataset}
\label{sec:dataset}
\begin{figure}
    \centering
    \includegraphics[width=1\linewidth]{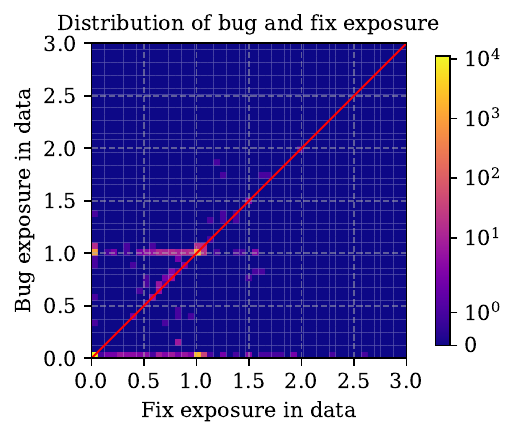} 
    \caption{Distribution of exposure of bugs and fixes in logarithmic scale}
    \label{fig:seenness}
\end{figure}

To understand the baseline exposure landscape, we first analysed the distribution of exposure across all SStuBs samples using Data Portraits membership inference on the Stack-v2 training corpus. \autoref{fig:seenness} illustrates the exposure distribution for both buggy and fixed code variants on a logarithmic scale. We observe that there are four major clusters, the first around a fix and bug exposure score of $0$, the second around exposure scores of $1$, a third around  $1$ and $0$ and a fourth around $0$ and $1$ respectively. 

There are a few samples with a fix exposure score less than $1$ and a bug exposure score of $1$. A score between $0$ and $1$ indicates that part of the sequence had a hit in the Data Portrait. But the length of the hit was not sufficient to conclude membership. Note that we only consider samples where the exposure is 90\% or more to be members, as per the original work~\cite{marone2023data}. 

Some have exposures higher than $1$, this can be caused by the bug or fix appearing multiple times in the training data. Some code snippets were duplicated dozens or even hundreds of times in the training data, likely due to repository forks, copy-pasted code, or independently written clones. 

As shown in \autoref{tab:seenness_tab}, the vast majority of samples ($11,286$ pairs, or 67\% of the dataset) had neither the bug nor the fix present in the training data. This finding suggests that most evaluation instances represent genuinely unseen code patterns. Among pairs with at least one variant exposed, fixes were substantially more likely to appear in the training corpus than bugs. Specifically, $2,335$ samples (14\% of the dataset) contained only the fix in Stack-v2, while only $1,109$ samples (7\%) contained exclusively the bug. A smaller subset of $2,169$ pairs (13\%) had fixed and buggy variants present in the training data. 

We examine the temporal dynamics of bug-fix pairs by analysing the number of commits between the introduction of a bug and its subsequent fix. As shown in \autoref{tab:seenness_tab}, the mean number of commits until fix varies substantially between exposure categories. Pairs where only the bug was seen exhibit the longest bug lifetime, with an average of 14,152 commits before correction, nearly double the dataset mean. In contrast, pairs where neither variant was seen have the shortest bug lifetime (4,735 commits on average). Pairs with both variants present show an intermediate lifetime (9,328 commits), while fix-only pairs average 6,373 commits before the fix.

\begin{table}[]
    \centering
    \caption{Count and mean number of commits until fix for each SStuB exposure category}
    \begin{tabular}{l|rr|r}
    Category        & Count & \#Commits & \%  \\
    \toprule
    Neither seen    & 11286 & 4735      & 67\%  \\
    Both seen       & 2169  & 9328      & 13\%  \\
    Only Bug        & 1109  & 14152     & 7\%   \\
    Only Fix        & 2335  & 6373      & 14\%  \\
    \toprule
    Total           & 16899 & 6169      & ~     \\
    \end{tabular}
    \label{tab:seenness_tab}
\end{table}

\begin{RQanswer}
Most of the bug-fix pairs in our data set were not present in the training corpus, and 67\% of the samples had neither variant seen. When exposure exists, fixes appear more frequently than bugs, and fixes tend to be longer on average than their buggy counterparts. Bug-only pairs exhibit nearly double the bug lifetime compared to the dataset mean, while neither-seen pairs show the shortest lifetimes.
\end{RQanswer}

\section{Results}
\label{results}
In this section, we present the empirical findings from our evaluation, addressing each research question in turn. For clarity and brevity, our analysis will primarily focus on the results of the StarCoder2-7B model. We found that the core patterns were remarkably consistent across the three StarCoder2 models evaluated. The complete results are available in the replication package.

\subsection{RQ1: Exposure Influence on Metrics}
\label{sec:rq1}

\begin{table*}[t]
    \centering
    \caption{Proportion of Preferred Variants by Bug xor Fix Exposure Across Models}
    \label{tab:xor_all}
    \begin{tabular}{l|rr|rr|rr|rr|rr|rr}
        Model & \multicolumn{4}{c|}{StarCoder2-7B} & \multicolumn{4}{c|}{Mellum-4B} & \multicolumn{4}{c}{SmolLM3} \\
        \midrule
        Exposure & \multicolumn{2}{c|}{Fix} & \multicolumn{2}{c|}{Bug} & \multicolumn{2}{c|}{Fix} & \multicolumn{2}{c|}{Bug} & \multicolumn{2}{c|}{Fix} & \multicolumn{2}{c}{Bug} \\
        \midrule
        Preferred & Fix & Bug & Fix & Bug & Fix & Bug & Fix & Bug & Fix & Bug & Fix & Bug \\
        \midrule
        length & 0.59 & 0.41 & 0.48 & 0.52 & 0.59 & 0.41 & 0.48 & 0.52 & 0.52 & 0.48 & 0.49 & 0.51 \\
        perplexity & 0.62 & 0.38 & 0.44 & 0.56 & 0.53 & 0.47 & 0.58 & 0.42 & 0.54 & 0.46 & 0.44 & 0.56 \\
        min_prob & 0.68 & 0.32 & 0.78 & 0.22 & 0.70 & 0.30 & 0.77 & 0.23 & 0.72 & 0.28 & 0.79 & 0.21 \\
        max_prob & 0.58 & 0.42 & 0.70 & 0.30 & 0.57 & 0.43 & 0.69 & 0.31 & 0.54 & 0.46 & 0.60 & 0.40 \\
        gini & 0.60 & 0.40 & 0.40 & 0.60 & 0.60 & 0.40 & 0.41 & 0.59 & 0.49 & 0.51 & 0.39 & 0.61 \\
        geometric_mean & 0.62 & 0.38 & 0.44 & 0.56 & 0.64 & 0.36 & 0.47 & 0.53 & 0.54 & 0.46 & 0.44 & 0.56 \\
        arithmetic_mean & 0.60 & 0.40 & 0.38 & 0.62 & 0.62 & 0.38 & 0.39 & 0.61 & 0.50 & 0.50 & 0.40 & 0.60 \\
    \end{tabular}
\end{table*}

\begin{table*}[t]
    \centering
    \caption{Proportion of Preferred Variants by Both and Neither Exposure Across Models}
    \label{tab:and_nor_all}
    \begin{tabular}{l|rr|rr|rr|rr|rr|rr}
        Model & \multicolumn{4}{c|}{StarCoder2-7B} & \multicolumn{4}{c|}{Mellum-4B} & \multicolumn{4}{c}{SmolLM3} \\
        \midrule
        Exposure & \multicolumn{2}{c|}{Both} & \multicolumn{2}{c|}{Neither} & \multicolumn{2}{c|}{Both} & \multicolumn{2}{c|}{Neither} & \multicolumn{2}{c|}{Both} & \multicolumn{2}{c}{Neither} \\
        \midrule
        Preferred & Fix & Bug & Fix & Bug & Fix & Bug & Fix & Bug & Fix & Bug & Fix & Bug \\
        \midrule
        length & 0.53 & 0.47 & 0.54 & 0.46 & 0.53 & 0.47 & 0.54 & 0.46 & 0.52 & 0.48 & 0.53 & 0.47 \\
        perplexity & 0.51 & 0.49 & 0.57 & 0.43 & 0.53 & 0.47 & 0.58 & 0.42 & 0.49 & 0.51 & 0.56 & 0.44 \\
        min_prob & 0.73 & 0.27 & 0.76 & 0.24 & 0.73 & 0.27 & 0.76 & 0.24 & 0.77 & 0.23 & 0.79 & 0.21 \\
        max_prob & 0.65 & 0.35 & 0.63 & 0.37 & 0.65 & 0.35 & 0.63 & 0.37 & 0.56 & 0.44 & 0.62 & 0.38 \\
        gini & 0.48 & 0.52 & 0.54 & 0.46 & 0.48 & 0.52 & 0.54 & 0.46 & 0.46 & 0.54 & 0.52 & 0.48 \\
        geometric_mean & 0.51 & 0.49 & 0.57 & 0.43 & 0.51 & 0.49 & 0.57 & 0.43 & 0.49 & 0.51 & 0.56 & 0.44 \\
        arithmetic_mean & 0.49 & 0.51 & 0.54 & 0.46 & 0.49 & 0.51 & 0.54 & 0.46 & 0.46 & 0.54 & 0.52 & 0.48 \\
    \end{tabular}
\end{table*}

In \autoref{sec:prefMetrics}, we introduced the metrics used in our study. As there is no known singular, ideal metric, in this section we evaluate metric suitability for selecting the fix over the bug. We study which metric reliably prefers fixes over bugs, \emph{and} remains stable when controlling for exposure. 

Table~\ref{tab:xor_all} present model preferences for pairs where exactly one variant (bug XOR fix) was seen during training. We pick an equal number of samples for each exposure setting (1109). This stratification isolates the effect of asymmetric exposure on model behaviour. When only the fix was present in the training data, the minimum probability showed the strongest bias towards correctness in all models, with 68–72\% of the pairs favouring the fix between different models. Maximum probability exhibited similar robustness. 

When only the bug was seen, minimum probability maintained its preference for fixes (77–79\% across models), resisting exposure bias. This suggests that minimum probability captures something fundamental about code correctness beyond simple pattern matching. Maximum probability also remained fix-preferring in bug-only scenarios (60–70\%), though with slightly reduced strength compared to fix-only cases.

In contrast, Gini coefficient and arithmetic mean showed exposure-driven preference reversal. When only the fix was seen, these metrics moderately preferred fixes (60–62\%), but when only the bug was seen, they flipped to preferring bugs (56–62\%). Length showed minimal discriminative power in either direction, hovering near 50-60\% across all conditions.

Perplexity reveals varying preferences. In fix-only conditions, all models favoured fixes. Under bug-only exposure, StarCoder2-7B and SmolLM3 shifted to prefer bugs, while Mellum-4B maintained a fix preference.

To further isolate learned correctness from exposure, \autoref{tab:and_nor_all} present results for pairs where both variants were seen (AND condition) or neither was seen (NOR condition). 

Across all three models, minimum probability maintained strong fix preference in both symmetric conditions (73–79\% for both-seen, 76–79\% for neither-seen). Gini coefficient and arithmetic mean lost nearly all discriminative power in symmetric conditions, hovering near or below 50\% fix preference. Geometric mean occupied a middle ground, showing moderate fix preference in neither-seen conditions but losing discriminative power in both-seen scenarios. Perplexity preferences were more balanced or fix-leaning. Both-seen scenarios show near parity. Neither-seen conditions consistently preferred fixes across models.

The preference patterns observed for StarCoder2-7B (\autoref{tab:xor_all}) replicate almost identically in Mellum-4B (\autoref{tab:and_nor_all}), despite Mellum being trained on additional proprietary data beyond Stack-v2. SmolLM3, though showing slightly weaker overall fix preference, exhibits the same relative ordering of metrics and the same exposure-driven reversals for Gini and arithmetic mean.

\begin{RQanswer}
Minimum probability and maximum probability consistently prefer fixes over bugs across all exposure conditions. In contrast, metrics like Gini coefficient and arithmetic mean show strong susceptibility to exposure bias, reversing their preference toward bugs when only the buggy variant is seen.
\end{RQanswer}

\subsection{RQ2: Exposure Effects by Bug Category}
\label{sec:rq2}
\begin{figure}
    \centering
    \includegraphics[width=\linewidth]{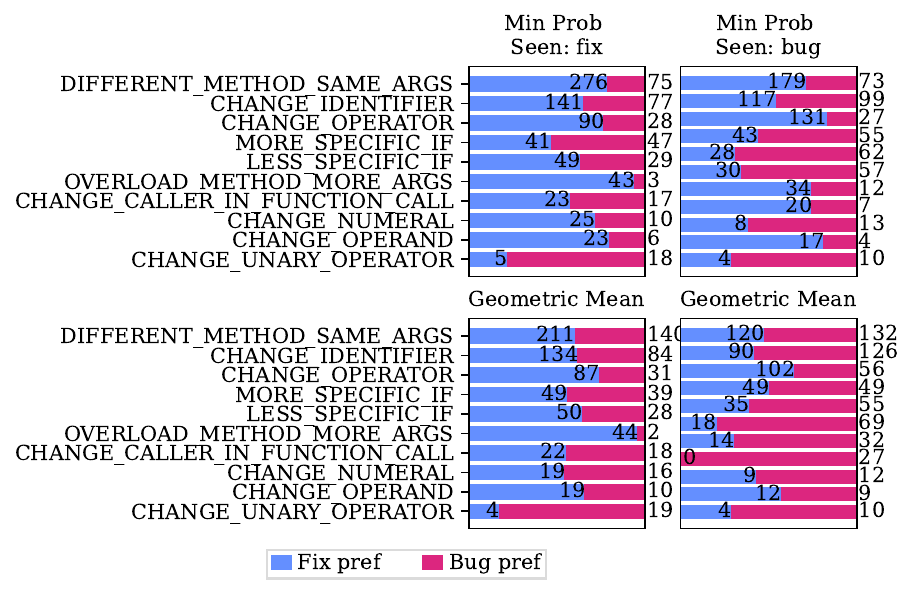}
    \caption{Model preference per bug category (StarCoder2-7B)}
    \label{fig:category_sc7b}
\end{figure}

\begin{figure}
    \centering
    \includegraphics[width=\linewidth]{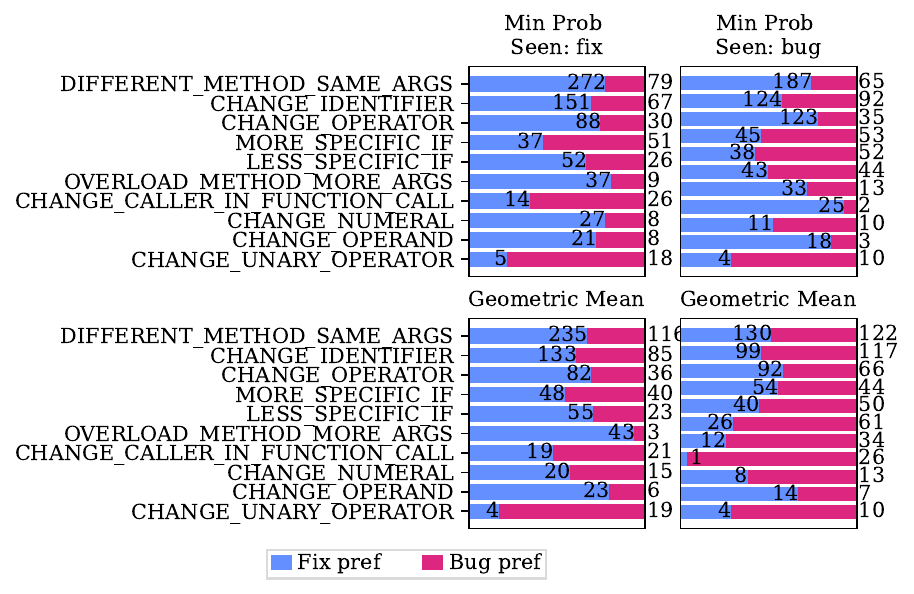}
    \caption{Model preference per bug category (Mellum-4B)}
    \label{fig:category_mellum}
\end{figure}

\begin{figure}
    \centering
    \includegraphics[width=\linewidth]{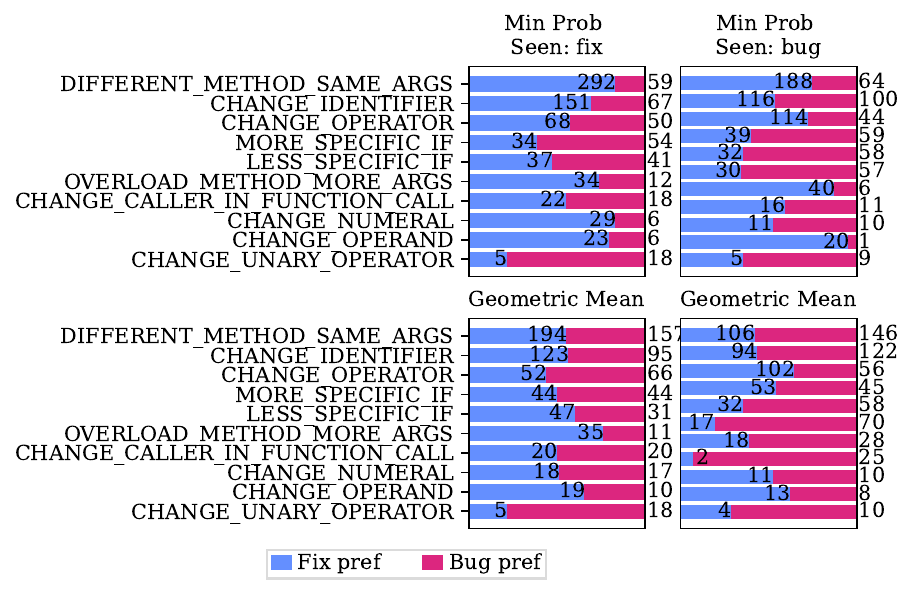}
    \caption{Model preference per bug category (SmolLM3)}
    \label{fig:category_smollm3}
\end{figure}

To investigate how bug-vs-fix preference varies by bug type, we stratified our analysis by the SStuBs bug categories for pairs where either the bug or the fix was seen. \autoref{fig:category_sc7b}, \ref{fig:category_mellum}, and \ref{fig:category_smollm3} show the preference breakdown for StarCoder2-7B, Mellum-4B, and SmolLM3, respectively. Each figure contrasts a metric which is biased towards fixes (min prob, top row) with an exposure-sensitive metric (geometric mean, bottom row) under two asymmetric exposure conditions: `Seen: fix' (left) and `Seen: bug' (right), note that these samples are exclusive, they belong to only one of the two exposure criteria, similar to \autoref{tab:xor_all}. The blue bars represent the proportion of pairs where the fix was preferred, while the red bars indicate a preference for the bug.

Across all models, the results confirm the findings from RQ1 at a more granular level: min prob consistently favours fixes even when only the bug was seen, whereas geometric mean often flips its preference to align with the exposed variant. For instance, in \autoref{fig:category_sc7b}, when only the bug is seen (top right), minprob still prefers the fix in most categories, such as DIFFERENT METHOD SAME ARGS (188 fix vs. 64 bug). In contrast, geometric mean's preference for many of these same categories shifts dramatically towards the bug (bottom right), as seen with CHANGE IDENTIFIER (94 fix vs. 122 bug).

Certain bug categories demonstrate strong robustness to exposure. For CHANGE OPERATOR, all models show a strong and stable preference for the fix, regardless of which variant was seen or which metric was used.

Conversely, some categories are susceptible to exposure-driven propagation. The CHANGENUMERAL category is a prime example. For all three models, both metrics show a preference flip. With Mellum-4B (\autoref{fig:category_mellum}), min prob preference shifts from 27-8 in favour of the fix to 11-10. The effect is even more pronounced for geometric mean, which shifts from 87-31 (fix) to 56-102 (bug). Similar flips are observed for LESS SPECIFIC IF and OVERLOAD METHOD MORE ARGS.

Finally, we identify categories where the model consistently prefers the bug, regardless of exposure. For CHANGE UNARY OPERATOR, all models show a strong preference for the buggy variant even when only the fix was seen. For Mellum-4B, the preference is 5-18 in favour of the bug. This suggests the `buggy' pattern may be far more common in the general training data or represents a strong prior that limited exposure to the fix cannot overcome. These patterns of category-specific robustness and susceptibility are consistent across all three models.

\begin{RQanswer}
Model preference varies significantly by bug category. Some categories are robust, with models consistently preferring the fix regardless of exposure. Others are more sensitive to exposure. A few categories, show an intrinsic model preference for the buggy version, which persists even when only the fix is seen.
\end{RQanswer}

\subsection{RQ3: Exposure Impact on Generations}
\begin{figure}
    \centering
    \includegraphics[width=0.8\linewidth]{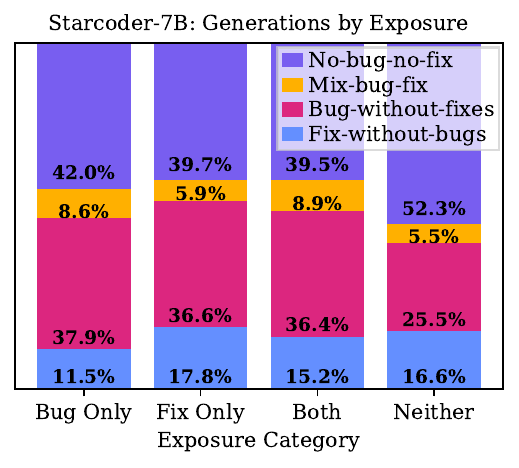}
    \caption{Model generations per exposure category (StarCoder2-7B)}
    \label{fig:gen_sc7}
\end{figure}

\begin{figure}
    \centering
    \includegraphics[width=0.8\linewidth]{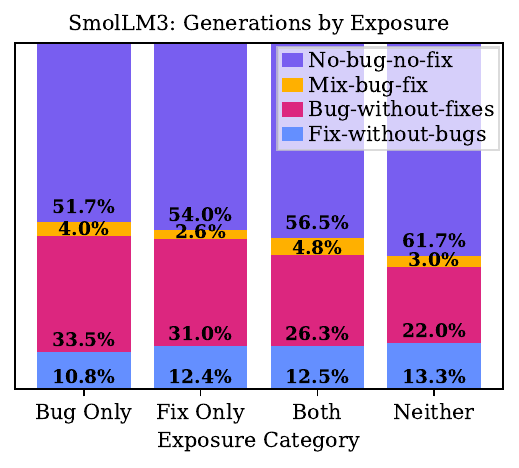}
    \caption{Model generations per exposure category (SmolLM3)}
    \label{fig:gen_smollm3}
\end{figure}

While RQ1 examines model preferences through likelihood-based metrics, RQ3 investigates actual generative behaviour by analysing what code models produce. We evaluate whether completions match the buggy variant, the fixed variant, both, or neither, stratifying results by exposure category to measure memorisation effects on generation.

The results for StarCoder2-7B (we omit Mellum-4B as the results were similar to the StarCoder2 models) and SmolLM3 are presented in \autoref{fig:gen_sc7} and \ref{fig:gen_smollm3}.

Across all models and exposure categories, generations match buggy code at substantially higher rates than fixed code, with bug-exclusive matches consistently outpacing fix-exclusive matches. For StarCoder2-7B (\autoref{fig:gen_sc7}), `Bug without fixes' rates were highest in bug-only exposure and lowest in neither-seen, while `Fix without bugs' rates were substantially lower across all categories. Mixed matches were moderate, and no-match rates were highest in the neither-seen category.

SmolLM3 (\autoref{fig:gen_smollm3}) shows lower overall match rates compared to StarCoder2, with similar patterns: `Bug without fixes' highest in bug-only and lowest in neither-seen, `Fix without bugs' consistently low, mixed matches minimal, and no-match rates predominant, especially in neither-seen.

The bug-only exposure category consistently showed the highest `Bug without fixes' rates across models. Conversely, fix-only exposure leads to modest increases in `Fix without bugs' for StarCoder2, but minimal or no increase for SmolLM3. The both-seen category exhibited high `Bug without fixes' rates for StarCoder2, comparable to bug-only, while SmolLM3's rates fall between bug-only and neither-seen.

\begin{RQanswer}
Models generate buggy variants at substantially higher rates than fixes across all exposure categories, with bug-exclusive matches consistently outpacing fix-exclusive ones. Bug-only exposure amplifies bug propagation the most, while neither-seen minimises it; fix-only provides minimal improvement in fix generation.
\end{RQanswer}

\section{Discussion}
\label{sec:discussion}
Our investigation reveals that code language models, despite encountering fewer bugs than fixes in their training data overall, exhibit a pronounced tendency to propagate bugs over fixes, heavily influenced by specific exposure patterns. Across all models, buggy variants were generated at substantially higher rates than fixes in every exposure category, with bug-only exposure exacerbating this bias. Addressing the core question of if ``models see, then do'': models only adhere to this principle to a certain extent. Exposure to bugs strongly drives increased bug generation and preference, whereas fix exposure only weakly promotes corrections. Metrics like minimum probability highlighted that models can generalise toward fixes in all exposure conditions.

The category-specific susceptibilities observed in RQ2, suggest that certain bug types align more closely with prevalent coding idioms or anti-patterns embedded in the training data. For instance, numeral changes may resemble common refactoring or optimisations that appear frequently in code repositories, making models more prone to propagating buggy variants when they are the only ones seen. This alignment could amplify security risks in LLM-generated code, as models might inadvertently reproduce vulnerable patterns. To mitigate this, targeted fine-tuning on under-represented fix patterns could bolster model robustness.

Another insight emerges from contrasting likelihood-based evaluation with actual generative behaviour. We observed that metrics like minimum token probability consistently favour the fixed code across all exposure conditions, whereas in practice the models often generate the buggy variant at much higher rates, especially under bug-only exposure. This reflects that these methods probe different aspects of the model. Likelihood metrics examine the probability distribution the model has learned, whereas generation metrics reflect the model's actual output under a given decoding scheme. In other words, the model ``knows'' that the fix should be more probable, but the process of sampling and decoding can still yield the bug as output. Both evaluation approaches can be complementary: likelihood-based metrics can diagnose what the model has learned and its confidence profile, whereas generation-based metrics reveal its practical behaviours and risks.

\paragraph{Bug Lifecycle}
Our results reveal that fixes appear more frequently in the training corpus than bugs, with 14\% of pairs having only the fix exposed, compared to 7\% exposing only the bug. This asymmetry may stem from the inherent construction of datasets like SStuBs, where bugs are mined from commit histories and can only be identified retrospectively once they are fixed. Bugs, by nature, have a limited observable lifetime in repositories, they persist until corrected, after which the fixed version dominates subsequent versions and forks. Consequently, training corpora like Stack-v2 are likely to capture more instances of the post-fix code, especially in actively maintained projects where fixes propagate quickly through merges and updates. 

Analysis of bug lifetimes reveals that detected bugs in exposure categories like bug-only pairs persist for substantially longer compared to fixes or neither-seen pairs. Cases where both variants are seen fall in between, suggesting a detection bias toward longer-lived bugs that survive multiple repository iterations before correction. Very short-lived bugs, which are quickly fixed within a few commits, may evade mining in datasets like SStuBs altogether. This could imply that our evaluation under-represents transient errors, where bugs have comparably short lifetimes, potentially skewing exposure-aware assessments toward more entrenched, harder-to-fix patterns that models are more likely to propagate. 

The observed preference for bugs in model generations (RQ3), even in scenarios where fixes are more exposed. Fix-only pairs show only marginal increases in fix-matching rates, this highlights a potential imbalance in training data where erroneous patterns are reinforced through sheer volume or recency bias in code repositories. Bugs, often introduced in recent commits before being fixed, may appear in more transient or duplicated contexts across forks and pull requests. In contrast, fixes, while more stable, might be diluted by the nature of LLM training, where high-frequency error patterns overshadow corrections.

Higher bug generation rates may also arise from biases in dataset mining and the inherent `naturalness' of human-written errors. SStuBs mining favours detectable bugs that were eventually fixed, potentially underrepresenting fixes that resemble common idioms, leading to an evaluation set where bugs appear more generation-prone. Additionally, these bugs were originally authored by programmers, making them feel `natural' in probabilistic terms; models may favour them as they align with frequent coding slips observed in training. Furthermore, these are coding slips that managed to get past unit-testing, and perhaps code review, into the code base. On the other hand, their eventual discovery and correction indicates they were detected, and fixed.

\paragraph{Models}
Interestingly, our analysis uncovered minimal between the 3B and 7B and 15B variants of StarCoder2, despite the substantial disparity in model size and capacity. The models exhibit comparable generation rates favouring bugs across exposure categories. Preference metrics like minimum probability further reinforced this uniformity. This lack of differentiation suggests that scaling model parameters alone may not suffice to overcome inherent biases in training data.

The near-identical performance of Mellum-4B and the StarCoder2 models across RQ1–RQ3, in contrast to the distinct patterns in SmolLM3, suggests that, for our evaluation, additional training data has minimal impact. Mellum-4B, while incorporating some closed-source datasets beyond Stack-v2, exhibits preference metrics and generation rates that mirror StarCoder2-7B closely, implying that these proprietary additions are limited in scope or relevance to code tasks. SmolLM3, however, deviates with lower overall match rates and weaker fix preferences, likely due to its significant inclusion of non-code training data; although code is not expected in these portions, incidental inclusions or domain shifts could dilute its code-specific capabilities.

From an AI safety perspective, the tendency of models to regenerate seen bugs, as evidenced by higher bug-matching rates in generations (RQ3), raises significant concerns about downstream harms, such as introducing vulnerabilities into production software. When models memorise and reproduce buggy patterns from training data, they risk propagating errors across multiple projects. This phenomenon aligns with broader discussions on memorisation risks in LLMs, where unintended data regurgitation can undermine software reliability~\cite{carlini2021extracting, alkaswan2024traces, jahanshahi2025cracks}.

While developers typically have little control over the composition of pre‑training datasets, our study highlights the specific risks that models may inherit from such data. Understanding these exposure effects is therefore essential, particularly because practitioners often deploy models without direct visibility into or control over their training sources~\cite{alkaswan2023abuse, wang2025comprehensive, alkaswan2025code}.

\subsection{Threats to Validity}
Although we have made efforts to ensure the robustness and reliability of our study, several potential threats to validity should be acknowledged. 

\paragraph{Internal Validity}
One threat is the potential for false positives in membership inference using Data Portraits, as Bloom filters can incorrectly indicate exposure due to hash collisions, with a reported false positive rate of 0.1\%~\cite{marone2023data}. This could lead to misclassification of bug-fix pairs into exposure categories, potentially exaggerating exposure-driven effects. We limited this impact by adhering to the tool's recommended query length threshold (99 tokens) and using the recommended 90\% badness score cutoff for binary classification~\cite{marone2023data}. 

A potential limitation in our analysis stems from the context provided to models, which may not be sufficiently constraining to unambiguously distinguish buggy from fixed code. We evaluate single-statement changes with the preceding file context, without full project context, meaning that the preceding lines alone might admit multiple valid completions. This could explain instances where generations match neither variant (prevalent in RQ3), as models generate plausible but unrelated code. 

Another threat arises from the probabilistic nature of LLM generations in RQ3, where decoding may introduce variability due to model stochasticity. We mitigated this by generating five independent completions per prompt and aggregating statistics (`Any bug-fix' rates), reducing the influence of single-run anomalies.

\paragraph{External Validity}
We rely on Java-specific bug-fix pairs from the SStuBs dataset, which may not generalise to other programming languages with different syntax or error patterns (such as dynamic languages like Python). This could limit the applicability of category-specific findings in RQ2 to multilingual settings. We limited this impact by selecting SStuBs for its focus on single-statement bugs, which are common across languages, and by using models trained on Stack-v2's diverse corpus (619 languages). However, we accepted the Java focus as a necessary scope limitation, given SStuBs' established use in code LLM evaluations~\cite{jesse2023large, karampatsis2020often, richter2022tssb}, and recommend future extensions to multilingual benchmarks.

We only evaluated a small set of open-weight models (StarCoder2 variants, Mellum-4B, SmolLM3), which may not represent closed-source models like GitHub Copilot or larger architectures. Differences in training data or scale could alter exposure effects. We mitigated this by including models of varying sizes (3B to 15B parameters). We did not evaluate the models using the fill-in-the-middle (FIM) objective. SmolLM3 does not support FIM~\cite{bakouch2025smollm3}, and StarCoder2-15B exhibits poor FIM performance due to known training errors by its authors~\cite{lozhkov2024starcoder2stackv2}.

\paragraph{Construct Validity}
Our binary classification of exposure based on Data Portraits oversimplifies memorisation, ignoring nuances like partial matches or contextual variations in training data. This could misrepresent true exposure, affecting downstream analyses. We accepted some simplification as a trade-off for scalability, given the size of Stack-v2.

Our analysis measures memorization using literal string matching, as partial or approximate matching lacks a well‑defined and reliable criterion for automated data extraction~\cite{hayes2025measuring}. Because buggy and fixed versions of code often differ by only a small number of tokens, generations that do not exactly match either version are difficult to classify, potentially leading to underestimation of partial memorization. We apply this strict criterion uniformly across all exposure categories, preserving the validity of relative comparisons. 

\subsection{Future Work}
Our exposure-aware evaluation framework opens several avenues for future research.

One promising direction is extending the analysis to multilingual and multi-dataset settings to enhance external validity. While our study focused on Java bug-fix pairs from SStuBs, incorporating benchmarks like Defects4J~\cite{just2014defects4j}, QuixBugs~\cite{lin2017quixbugs, prenner2022can} or PySStuBS~\cite{kamienski2021pysstubs}, which span languages such as Python and C++, could reveal language-specific exposure effects. For instance, dynamic languages might exhibit different bug lifetimes or category susceptibilities (such as type-related errors in RQ3), potentially modulated by syntactic differences. By applying Data Portraits to the full Stack-v2 corpus, future work could quantify cross-language propagation risks, providing a more comprehensive view of how code LLMs handle diverse error patterns.

Improving membership inference techniques represents another key area, and would solve the internal validity threats from false positives and binary classifications in Data Portraits. Advanced methods, such as embedding-based similarity searches or neural membership inference~\cite{shokri2017membership, hu2022membership, mireshghallah2022quantifying, salem_ml-leaks_2018}, could offer finer-grained exposure estimates, accounting for partial matches or contextual variations in training data. This would enable nuanced stratifications beyond our seen/unseen dichotomy, such as weighting by exposure frequency or recency, and better isolate memorisation from generalisation in preference metrics. Black-box membership inference could allow for the detection of exposure, purely based on model characteristics. However, current model-only membership inference approaches are not reliable enough to assist in our evaluation~\cite{meeus2025sok, dubinski2023towards}.

Finally, the metrics discussed in RQ1 and RQ2 might offer practical utility for developers by helping rank and filter multiple code completion candidates to select the most reliable option. Furthermore, metrics like the Gini coefficient can function as a form of membership inference, allowing developers to identify, and decide what to do with, code that has been memorized directly from the training data, without access to the training corpus.

\section{Conclusion}
In this paper, we explored whether code LLMs ``see, then do'' by propagating bugs from training data, hypothesising that exposure to buggy versus fixed code could drive memorisation and error regurgitation. Through an exposure-aware pipeline combining membership inference, likelihood metrics, and generation analysis on the SStuBs benchmark, we empirically examined this across multiple models.

Our results confirm that models do ``see, then do'' bugs: while many bug-fix pairs remain unseen and fixes are generally more exposed than bugs, models often prefer and generate buggy variants, especially when exposed to them. Some likelihood preferences lean toward fixes but can reverse under bug exposure, with effects varying by bug type.

Our study underscores the need for exposure-aware evaluation in coding LLMs. The ``see, then do'' effect reveals that models internalise fixes as preferable in likelihood scores, yet frequently reproduce memorised bugs in generations. This discrepancy between knowledge and outputs has implications for tool reliability, leading us to advocate membership testing and diverse metrics to uncover memorisation biases. By controlling for exposure, researchers can better assess true generalisation, mitigate error propagation, and promote safer software development.


\newpage
\bibliographystyle{ACM-Reference-Format}
\bibliography{references.bib}
\end{document}